\documentclass[preprint,12pt]{elsarticle}

\usepackage{hyperref}
\usepackage{multicol}
\usepackage{color}
\usepackage{xspace}
\usepackage{pdfwidgets}
\usepackage{enumerate}
\usepackage{tabularx}
\usepackage{subcaption}
\usepackage{lineno}

\def\ttdefault{cmtt}

\usepackage{multirow}

\journal{}

\begin{document}

\pinclude{\copy\contbox\printSq{\LastPage}}

\begin{frontmatter}

\title{The Control Unit of the KM3NeT Data Acquisition System}

\cortext[cor]{corresponding author}

\author[1]{S.~Aiello}
\author[2]{F.~Ameli}
\author[3]{M.~Andre}
\author[4]{G.~Androulakis}
\author[5]{M.~Anghinolfi}
\author[6]{G.~Anton}
\author[7]{M.~Ardid}
\author[8]{J.~Aublin}
\author[4]{C.~Bagatelas}
\author[9,10]{G.~Barbarino}
% ...
\address[1]{INFN, Sezione di Catania, Via Santa Sofia 64, Catania, 95123 Italy}
\address[2]{INFN, Sezione di Roma, Piazzale Aldo Moro 2, Roma, 00185 Italy}
\address[3]{Universitat Polit{\`e}cnica de Catalunya, Laboratori d'Aplicacions Bioac{\'u}stiques, Centre Tecnol{\`o}gic de Vilanova i la Geltr{\'u}, Avda. Rambla Exposici{\'o}, s/n, Vilanova i la Geltr{\'u}, 08800 Spain}
\address[4]{NCSR Demokritos, Institute of Nuclear and Particle Physics, Ag. Paraskevi Attikis, Athens, 15310 Greece}
\address[5]{INFN, Sezione di Genova, Via Dodecaneso 33, Genova, 16146 Italy}
\address[6]{Friedrich-Alexander-Universit{\"a}t Erlangen-N{\"u}rnberg, Erlangen Centre for Astroparticle Physics, Erwin-Rommel-Stra{\ss}e 1, 91058 Erlangen, Germany}
\address[7]{Universitat Polit{\`e}cnica de Val{\`e}ncia, Instituto de Investigaci{\'o}n para la Gesti{\'o}n Integrada de las Zonas Costeras, C/ Paranimf, 1, Gandia, 46730 Spain}
\address[8]{APC, Universit{\'e} Paris Diderot, CNRS/IN2P3, CEA/IRFU, Observatoire de Paris, Sorbonne Paris Cit\'e, 75205 Paris, France}
\address[9]{INFN, Sezione di Napoli, Complesso Universitario di Monte S. Angelo, Via Cintia ed. G, Napoli, 80126 Italy}
\address[10]{Universit{\`a} di Napoli ``Federico II'', Dip. Scienze Fisiche ``E. Pancini'', Complesso Universitario di Monte S. Angelo, Via Cintia ed. G, Napoli, 80126 Italy}
% ...

\date{\today}

\begin{abstract}
The KM3NeT Collaboration runs a multi-site neutrino observatory in the Mediterranean Sea. 
Water Cherenkov particle detectors, deep in the sea and far off the coasts of France and Italy, are already taking data while incremental construction progresses. Data Acquisition Control software is operating off-shore detectors as well as testing and qualification stations for their components.
The software, named \textit{Control Unit}, is highly modular. It can undergo upgrades and reconfiguration with the acquisition running. Interplay with the central database of the Collaboration is obtained in a way that allows for data taking even if Internet links fail. In order to simplify the management of computing resources in the long term, and to cope with possible hardware failures of one or more computers, the KM3NeT Control Unit software features a custom dynamic resource provisioning and failover technology, which is especially important for ensuring continuity in case of rare transient events in multi-messenger astronomy. The software architecture relies on ubiquitous tools and broadly adopted technologies and has been successfully tested on several operating systems. 
\end{abstract}
\end{frontmatter}

\section{Introduction}

The KM3NeT neutrino detectors are complex objects \cite{loi20} designed for neutrino astrophysics \cite{pointsrc} and the study of atmospheric neutrino oscillations \cite{limres}. They are being built an the bottom of the Mediterranean Sea in a phased installation scheme. The infrastructure will consist consist of three-dimensional arrays of photosensors also called \emph{building blocks}. Each building block will comprise 115 vertical instrumented detection lines (Detection Unit, DU) equipped with 18 optical sensors (Digital Optical Module, DOM). Each DOM \cite{dom} contains 31 photo-multiplier tubes (PMTs) that detect the Cherenkov light induced by relativistic particles emerging from neutrino interactions. The French site will host one such building block (Oscillation Research with Cosmics in the Abyss - ORCA) and the Italian site will host two building blocks (Astroparticle Research with Cosmics in the Abyss - ARCA). In each DOM, the data recorded by the PMTs are digitised and transferred to the shore station by the Central Logic Board (CLB).
The settings and performance of PMTs need to be controlled and monitored. Prototype DOM \cite{protodom} and DU \cite{protodu} have been successfully operated in the sea before deploying and running detection units according to the final design. The DOMs host also other instruments devoted to monitoring and dynamic position reconstruction, as the detector shape in water currents is constantly changing. In particular, an acoustic positioning system is in place taking data from hydrophones that listen to known emitters. At the base of each DU there is a module that contains some instruments and a CLB. The Trigger and Data Acquisition System (TriDAS) \cite{tridasvlvnt2018} relies on a distributed and scalable architecture. The computing processes that implement the TriDAS have a number of running instances that may grow as needed, exceeding a few hundreds on tens of servers in a single installation. Each detector can run different tasks, with varying data taking strategies. The Control Unit (CU), a suite of computer processes exposing distributed services, has the task of directing all such hardware and software components to work together. The Control Unit is also in charge of collecting and storing logs of operations that are suitable both for machine processing and human access.

In addition to the above stated needs, the qualification and certification procedures for single PMTs, DOMs or whole DUs require running one or more data acquisition tasks in controlled environments and with multiple testing protocols \cite{pmtchar} to ensure that all devices operate within specifications. The software running in detector operation is also used for production and testing of components. Test bench stations \cite{pmttest} actually work in a way that is very similar to shore stations of detectors for physics data taking.

KM3NeT searches for rare events - interactions both of primary cosmic neutrinos and of secondary neutrinos from cosmic rays - that may occur at any time. Maximising the detector livetime is a key requirement to collect high statistics. Hence the reliability of the Control Unit and the possibility to operate continuously despite hardware or software failures have a direct impact on the statistical significance of data taking results. 

The detectors are designed to operate at least for 10 years in the sea. The software makes use of widely adopted standards (see ahead in the text) at its foundation, with a large development and user base that should ensure support for a long time scale. All custom code is completely under the control of the KM3NeT Collaboration, whose software quality plan includes long-term software preservation.

After an overview in Section \ref{Overview} of the distributed architecture, the present paper describes the various services it consists of. Authentication and identification of users and services are described in Section \ref{auth}. Run control and overall supervision are described in Section \ref{MCP}. Details about the representation of detectors and operational parameters in the database are given in Section \ref{DetRS}. Interaction with the database is described in Section \ref{DBI}. Control of detector devices and instruments is described in Section \ref{DM}. In Section \ref{TM} it is shown how the software components of the data processing chain are controlled. Details of the networking protocols and services are given in Section \ref{Networking}. Dynamic resource provisioning and fault tolerance are described in Section \ref{DRP}. Conclusions are given in Section \ref{Conclusions}. For convenience, all acronyms are listed in the Appendix.

\section{Software components} \label{Overview}

The Control Unit consists of five different services that can run independently of each other:

\begin{enumerate}
    \item Local Authentication Provider (LAP);
    \item Master Control Program (MCP);
    \item Database Interface (DBI);
    \item Detector Manager (DM);
    \item TriDAS Manager (TM).
\end{enumerate}

The services can run on the same machine or on different servers, in the case of installations with failover functions. All programs are written in C\#, as specified in the standards ECMA-334:2003-2006, ISO/IEC 23270:2003-2006 and following. The executables are encoded in a machine-independent language that are JIT-compiled by the Mono \footnote{\label{mono} \url{https://www.mono-project.com}} compiler and can then run on different operating systems such as flavours of GNU/Linux \footnote{\label{linuxdistros} The distributions tested include Fedora 24, SLC6, CentOS 7, Debian 8 ``Jessie'', Debian 9 ``Stretch'', LMDE 2, Linux Mint 19 and Ubuntu 16.04 LTS, but there are no evident reasons for incompatibility on others.}, Microsoft Windows \footnote{\label{windowsos} Windows 7 / 2008 or higher, all desktop and server versions.} and OS X. In the KM3NeT context, the Control Unit is hosted by servers running CentOS 7 or SLC6.

\begin{figure}[ht!]
\includegraphics[width=0.9\textwidth]{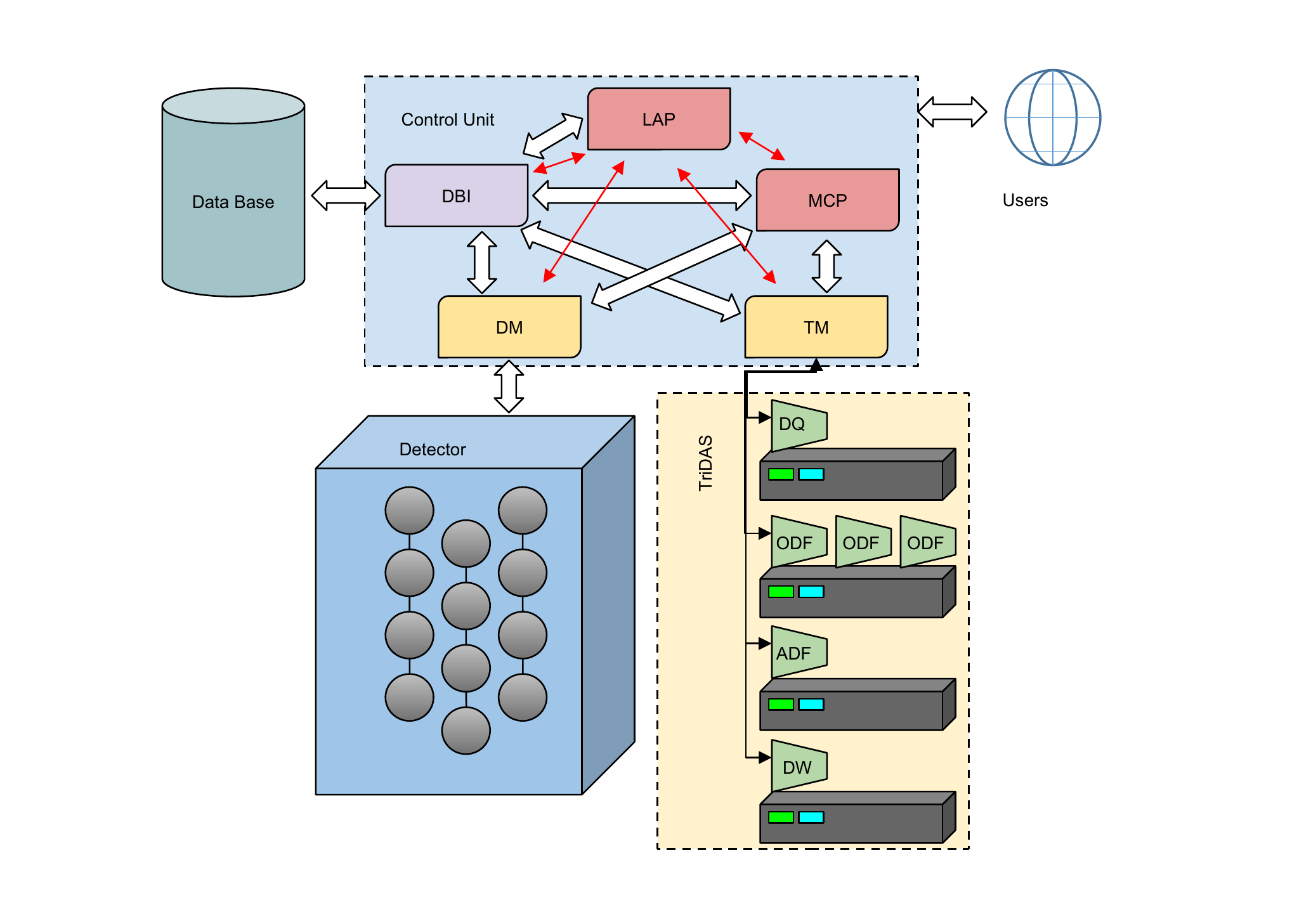}
\caption{\textit{The Control Unit components and their relationships. White and black arrows represent flows of control and monitoring information. Red arrows show the flow of authentication / authorization information. The flow of PMT and acoustic data from the detector to the TriDAS and hence to the final storage is not shown.}}
\label{fig-overview}
\end{figure}

All services have been developed to have a small footprint in terms of CPU and memory usage. They may run in more than one process on different machines for failover purposes (see Sect. \ref{DRP}) or to meet high demands in terms of workload. The latter case is foreseen for DM controlling large detectors, e.g. the full ARCA installation with a total such as 230 DUs.

Each service has a unique access point through HTTP \footnote{\label{HTTPS}Secure communication on SSL / TLS could also be supported, but in a local private network of the KM3NeT ICT infrastructure this is overkill.}. The Graphical User Interface (GUI), when present, is offered as a Web-like service on the HTTP. The GUI can be accessed by HTTP on VPN (Virtual Private Network) from remote controllers that pass both VPN authentication and CU authentication (see Sect. \ref{auth}). For highly critical management purposes and basic configuration, a local console accessible only by administrators through terminal is provided. The risk of misconfiguration is assessed to be higher if coming from inexperienced users than from remote attackers.
Figure~\ref{fig-overview} shows the logical connections among the services and with the detector and TriDAS components. In addition to control and logging, the Control Unit is also the bridge between the users, the central KM3NeT database \cite{dbvlvnt2015}, and the off-shore detector and the online trigger system. 

\section{Authentication and identification - LAP}
\label{auth}

Access to detector control and management is given to users on the basis of an authentication system, which is managed by the Local Authentication Provider. The LAP uses accounts and session tokens to manage identification and authentication. All accounts are kept in an encrypted local file together with the security credentials and privileges. When a login request is accepted for an account, the corresponding privileges are copied to a new session token that is then kept active until it expires or is deleted because of an explicit logout.
The LAP uses the logical scheme of account management shown in Table \ref{acctypes}.

\begin{table}[ht!]
    \begin{tabularx}{\linewidth}{l l X}
    \hline\\
    \multirow{2}{*}{User account}& Local account & The unique identifier, name and password are created locally in the detector / test bench control station and are meaningless outside of it \\
    \cline{2-3}\\
    & Global account & The unique identifier, name and password are managed on the central database of KM3NeT and are periodically synchronized with a local encrypted cache \\
    \hline\\
    \multicolumn{2}{l}{Service account}  & Defines a common name for a CU service\\
    \hline\\
    \end{tabularx}
    \caption{\textit{Account types.}}
    \label{acctypes}
\end{table}

Operating privileges are given to user and service accounts to enable specific functions such as controlling the whole station in terms of jobs (high level) or tuning single parameters (low level).
It is worth noticing that the function of a service depends on its privileges rather than on its name. This allows flexibility in the design: in the future, a single process may incorporate more than one function  and this would just need a change in the registration on the LAP rather than statically hard-coding an association between a name and a function.

A user can be granted privileges one by one or in well-defined groups named roles. Because detectors take data 365 days a year, 24 hours a day, the KM3NeT Collaboration follows a shift plan to share the load of detector control. Each shift lasts seven days and a shift team includes a \emph{Shifter} and a \emph{Shift Leader}, with the tasks of monitoring the detector operation and checking data quality. The \emph{Run Coordinator} stays in charge for a longer timespan (usually four-eight weeks), connecting the activity of each shift team to the next and overseeing the optimisation of the detector performance. The role system is especially useful for shift management: when a user is registered on the central database for a shift, he / she gets automatically and for the corresponding time window all the privileges that are defined in the \emph{Shifter} / \emph{Shift Leader} / \emph{Run Coordinator} role. They are all revoked when the shift ends. A user that is registered as a \emph{DAQ Expert} (Data AcQuisition expert, usually among the lead developers of hardware or software components) or \emph{Detector Operation Manager} (responsible for detector management, usually for several years) on the central database automatically gets all the related privileges on all installations. For example, shifters are supposed to operate the detectors using predefined configurations, whereas experts are allowed to tune single parameters for diagnostic and testing purposes.

While the concept of a user login is quite intuitive, a service login deserves some explanation. The mere fact that a program is installed and running on a server is not enough for it to be known to the LAP (and hence to other CU services). When the program logs in on the LAP it gets its own security token and becomes known to all other CU services. This explicit login requirement ensures that the hardware resource usage can be optimised and services can be moved from one machine to another according to the needs. This also makes the initial configuration easier, because there is no need to handcraft a static configuration file. Administrators can build the configuration indirectly by issuing incremental commands to the LAPs to register new instances of services.

\section{Run control - MCP} \label{MCP}

\begin{figure}[ht!]
\includegraphics[width=0.9\textwidth]{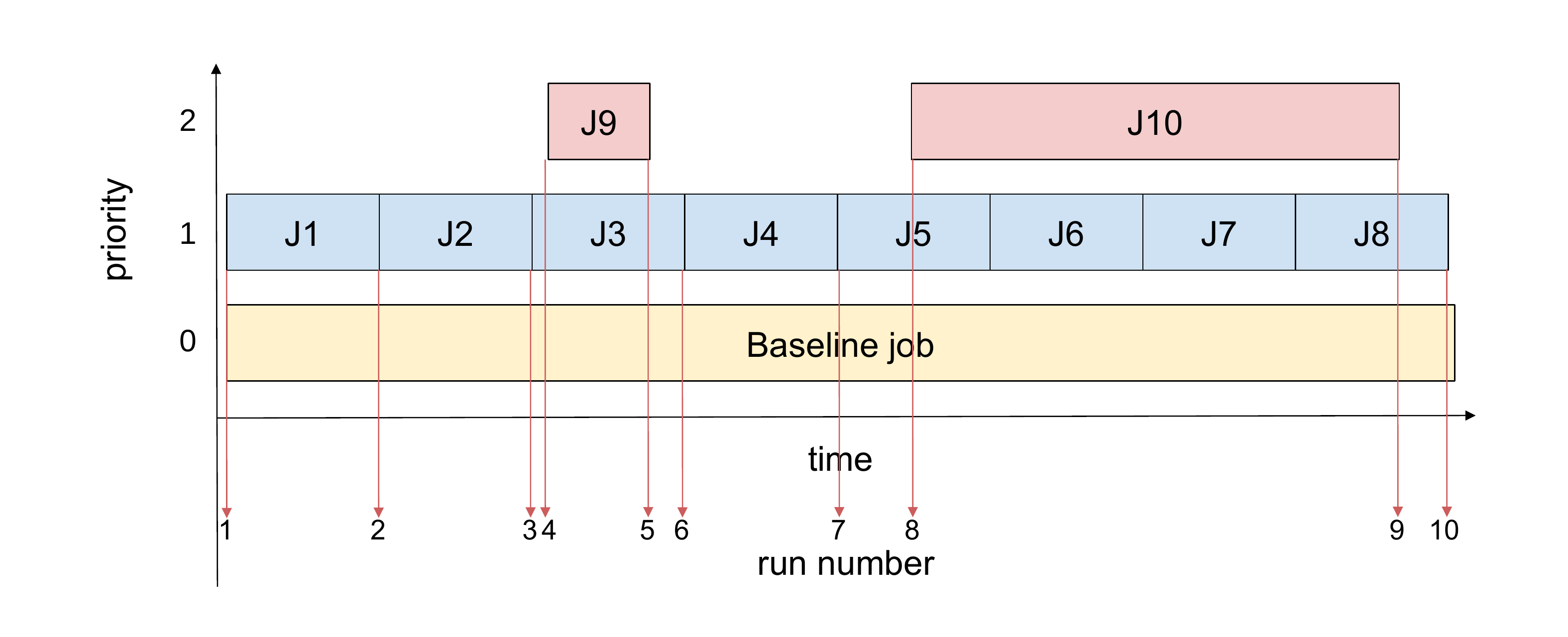}
\caption{\textit{Example of job chart. See explanations in the text.}}
\label{fig-jobs}
\end{figure}

The Master Control Program is in charge of maintaining the run status of the detector and TriDAS. The complete information of the run status consists of the following pieces:

\begin{enumerate}
\item Current detector: a detector changes when DUs are added or removed or for a failover reconfiguration (see Sect. \ref{DRP}).
\item Current \emph{runsetup}: the coherent set of input parameter values  controlling the detector, such as PMT supply voltage, and quantities to be read out for logging. See Sect. \ref{DM} and \ref{TM} for more details.
\item Current run number: a run is a timespan during which a detector is operated with the same \emph{runsetup}; for practical reasons a long run may be split in two or more with the same \emph{runsetup} to have smaller output datafiles.
\item Current target: the overall target of the CU can be one of the following (notice that a target change does not imply a run switch):
    \begin{itemize}
        \item Off: all PMTs are turned off, data taking is off, no triggering or post-processing.
        \item On: all PMTs are on, data taking is off, no triggering or post-processing.
        \item Run: all PMTs are on, data taking is active, triggering and post-processing run.
    \end{itemize}
\item Current time / position calibration: the set of adjusted positions and time offsets for individual PMTs that is being used for online triggering. 
\item Current job: a job is a run schedule with a priority grade. A run may start with or without a predefined schedule because the MCP may be commanded to immediately switch the run number. A job is a promise that at some time a new run will start with a \emph{runsetup} that is defined in advance and that will last for a certain timespan, unless preempted by higher priority jobs. One job may correspond to one or more runs. Some examples on job management are shown in Figure \ref{fig-jobs}: the baseline job is usually defined to use a \emph{runsetup} with tuned PMT voltages and the detector in ``On'' state; jobs J1-J8 might be routine data taking jobs with priority 1 and the detector in Run state; job J9 might be a calibration run and job J10 might be running a special data taking. Routine jobs J6 and J7 will produce no runs because they will be overridden by J10. J3 will produce two runs because the MCP will start with it, switch to J9 after J3 has started and then fall back to J3 again when J9 ends.
\end{enumerate}

Jobs may be modified before they begin and they can be truncated when they have started. The run status, run switch history and job addition / deletion / modification history are kept in a dedicated local file, which acts as a transaction log. Such information is periodically pulled by the DBI to be recorded in the central database \footnote{\label{dbccin2p3} Remotely hosted in the computing centre CC-IN2P3 in Lyon - \url{https://cc.in2p3.fr}.}. Only after the information has been successfully written in the database, the file is purged. In addition, all run switches are recorded to a human-readable log file, but the syntax is such that, in case of loss or corruption of the run status file, it is possible to reconstruct the latter from the former.

In standard operation, a detector may be required to run for months with the same operating parameters. For this purpose, it is possible to use the ``auto-schedule'' feature that automatically fills a priority line with jobs of equal duration and a specified \emph{runsetup} and target. This frees shifters from error-prone repetitive tasks.

Whenever the run state changes, the MCP notifies all the services that are registered in the LAP with the \emph{Status\_Notification\_Privilege}) privilege, which usually means at least DM and TM. This is a ``push'' type notification, aimed at fast communication. Fault tolerance is ensured by the ``pull'' communication mode: the DM and TM periodically update their knowledge of the run state by retrieving such information from the MCP. A finite time will elapse between the run switch by the MCP and the reaction in the DM and TM. All this is logged and it is possible to precisely identify the run switch latency time in each case.

A run switch is also triggered by a system reconfiguration after a fault (more detail in Sect. \ref{DRP}). It is worth pointing out that the MCP alone is responsible for providing a unique pair of detector and run number for each run. Different detectors in different KM3NeT sites can use the same run number without clashes.

\begin{figure}[ht!]
\centering{\fbox{\includegraphics[width=0.9\textwidth]{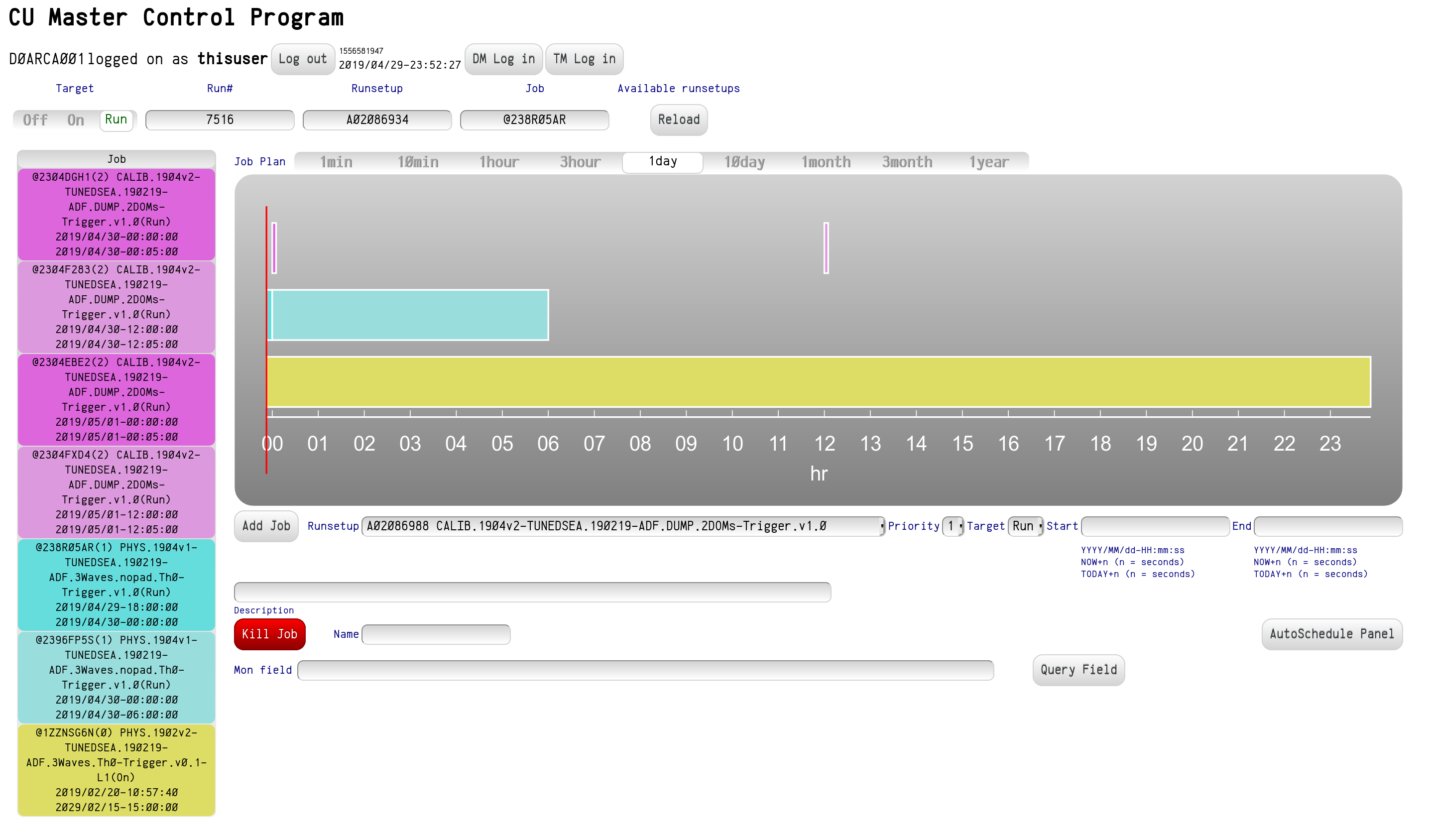}}}
\caption{\textit{Screenshot of the MCP graphical user interface, with several jobs scheduled at different priorities. }}
\label{fig-mcp-screen}
\end{figure}

The MCP offers a Web-based GUI (Figure \ref{fig-mcp-screen}) to perform all routine tasks, with the exception of service configuration and disaster recovery. The GUI enforces user privilege compliance: job scheduling is not allowed to users that are neither on shift nor \emph{DAQ Expert} privilege owners. An additional security check layer involving LAP queries is able to filter out possible HTTP forged queries that may try to circumvent or bypass the GUI. In this context, HTTPS would be possible but overkill because security is focused on ensuring compliance, by users and automated processes, to data taking procedures. All communications already occur on a private network and users connect through a VPN.

\section{Detector description and \emph{runsetups}} \label{DetRS}

\subsection{Detectors}

KM3NeT detectors are described and defined in the central database. A detector always has a location and a start timestamp, which is the first time it is connected and can provide signals. The end timestamp is set on its final disconnection. The same physical detector, located in the same place and reconnected, would have a different detector identifier. From the point of view of the CU, a detector is a list of previously integrated DUs and TriDAS processes, namely Data Queues (DQs) to rearrange data packets from single DOMs into events, Optical / Acoustic Data Filters (ODFs / ADFs) to run triggering algorithms and Data Writers (DWs) to write data to disk. In a basic implementation, TriDAS processes are defined in the central database, the process map is static and there can be no failover plan. In a more evolved view that supports dynamic provisioning and failover, the set of TriDAS processes might change during the lifetime of a detector and even several times per day in case of failures or addition of computing power. Nevertheless, from the point of view of MCP, DM and TM, there is only one definition of a detector that is provided at a certain time, and it always includes TriDAS processes.

\subsection{\emph{Runsetups}}

PMTs need their operating high voltage (HV) to be fine-tuned in order to provide uniform performance. The optimal value might also change over time. Likewise, functions may need to be enabled or disabled on certain DOMs, especially for testing and calibration purposes. \emph{Runsetups} define the input to each DOM and the output for feedback, monitoring and data logging, and all of them depend on the purpose each \emph{runsetup} was defined for (e.g. minimum data filtering, timing tuning, HV tuning, etc.). Many \emph{runsetups} differ only for some sets of parameters. Parameters with correlated meanings and purposes (e.g. PMT HVs, threshold and activation state) coalesce into \emph{configuration groups}. Each \emph{runsetup} is an ordered list of configuration groups, which are picked at various levels as referring to a whole category of items, subcategory or individual items.

\section{Interaction with the database - DBI}
\label{DBI}

\begin{figure}[ht!]
\includegraphics[width=0.9\textwidth]{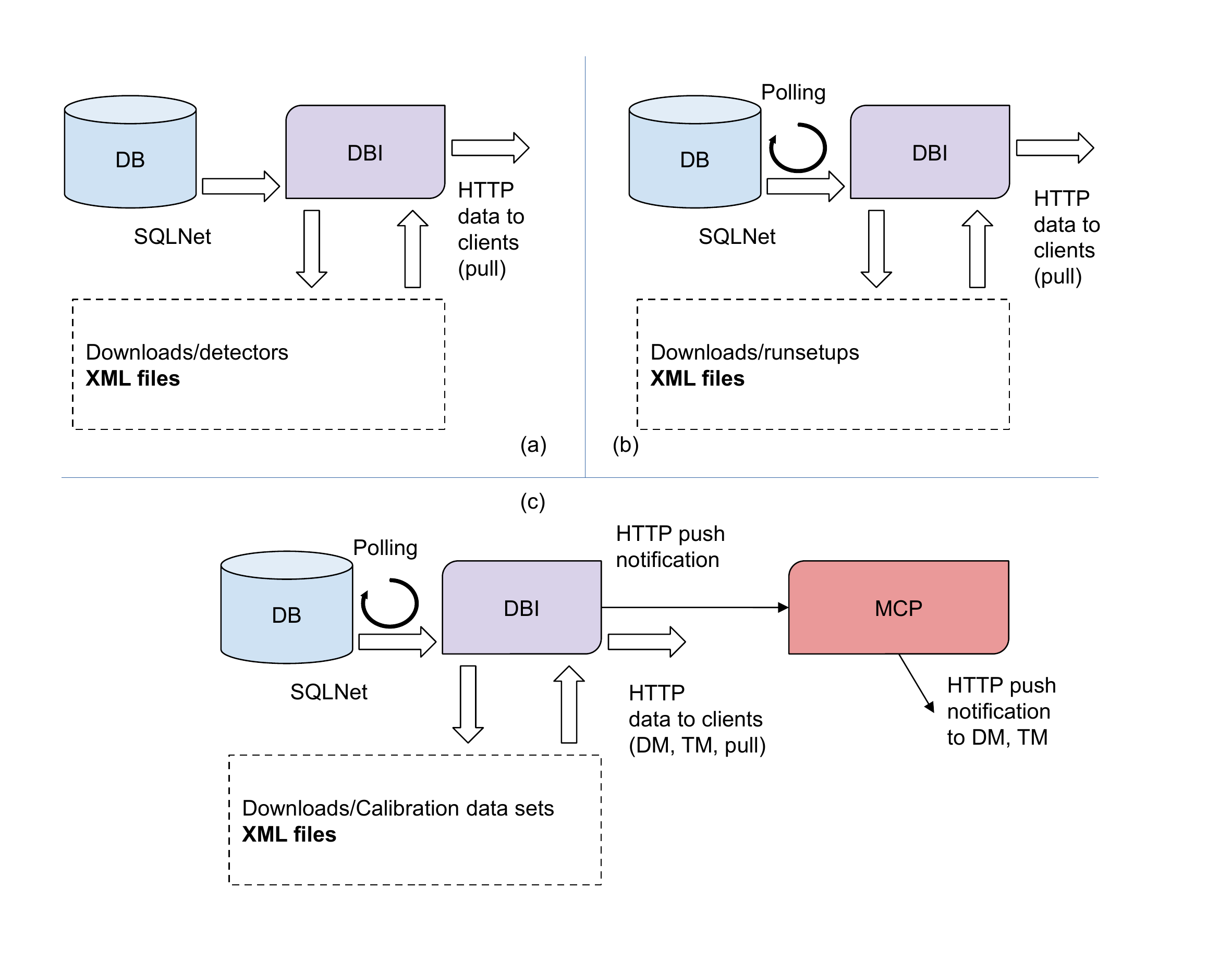}
\caption{\textit{Logic and network protocols involved in data download from the database. (a) Detector data flow. (b) Runsetup data flow. (c) Calibration data flow.}}
\label{fig-dbi}
\end{figure}

The service named Database Interface (DBI) is devoted to handling the interaction with the central database. Its main operating principle is to work as a file buffer to replace SQL / DML interaction of programs with the database as sketched in Figure \ref{fig-dbi}. The main reasons to implement a DBI are:

\begin{itemize}
    \item To avoid redundancy, database access credentials are stored in a single place at CU installation time and encrypted for safety.
    \item Decoupling CU code and database code / schema. SQL queries and / or DML statements need not be written in any code outside of the DBI itself. All the complications of handling and converting database data types are handled by the DBI and the client code is written in terms of CU data structures. This allows for refactoring on either side, i.e. the necessary evolution over time of both the CU and database will not affect each other.
    \item Coping with remote connection instability. The connection with the central database uses a Wide Area Network, which is intrinsically unreliable. The DBI stores all the datasets that are needed for CU operation in a local cache, speeding up access and improving reliability. On the other hand, the DBI buffers write operations and replays them if they fail because the database is not accessible.
\end{itemize}

Information sets that have been downloaded from the database are hosted in the download cache in XML format. As shown in Figure \ref{fig-dbi}, they include:

\begin{itemize}
    \item The current detector definition. It is downloaded only when the detector definition changes, by authorized users.
    \item All \emph{runsetups} written for the current detector (a one-to-many relationship). The DBI regularly polls the database for appearance of new \emph{runsetups}, but on-demand access is tried for \emph{runsetups} required by the MCP / DM / TM that are not yet in the cache.
    \item The current sets of calibration data. These data are continuously polled for updated versions and immediately pushed to the MCP and other services.
\end{itemize}

While \emph{runsetups} are usually created by humans and the time of their creation is well separated from the time they are used, calibration datasets are supposed to be updated regularly and automatically to have optimal detector operation. As soon as a new set is available and has been successfully downloaded, the DBI notifies the MCP which decides when to switch the run and broadcasts the signal to other services. In this sense, the DBI is an active part in data taking. 

The upload cache stores data queued to be written in the database, usually flushed upon successful transfer. In this case, binary files are expected in the native format generated by writer programs. The DBI handles the needed conversions. At present, the following types of data are hosted in the upload cache:

\begin{itemize}
    \item DM \emph{datalogs} that contain detector monitoring data and notification of management events, such as the real time of run start for each CLB, which is different from the time the MCP issues the command to change the run number; see Sect. \ref{DM} for more details.
    \item TM \emph{datalogs} containing logs of TriDAS activity, documenting actual starting-stop times of each run process-by-process, possible crashes, etc.; see Sect. \ref{TM} for more details.
    \item Times-Of-Arrival (TOAs) of acoustic wave pulses found by the ADF(s).
\end{itemize}

\begin{figure}[ht!]
\includegraphics[width=0.9\textwidth]{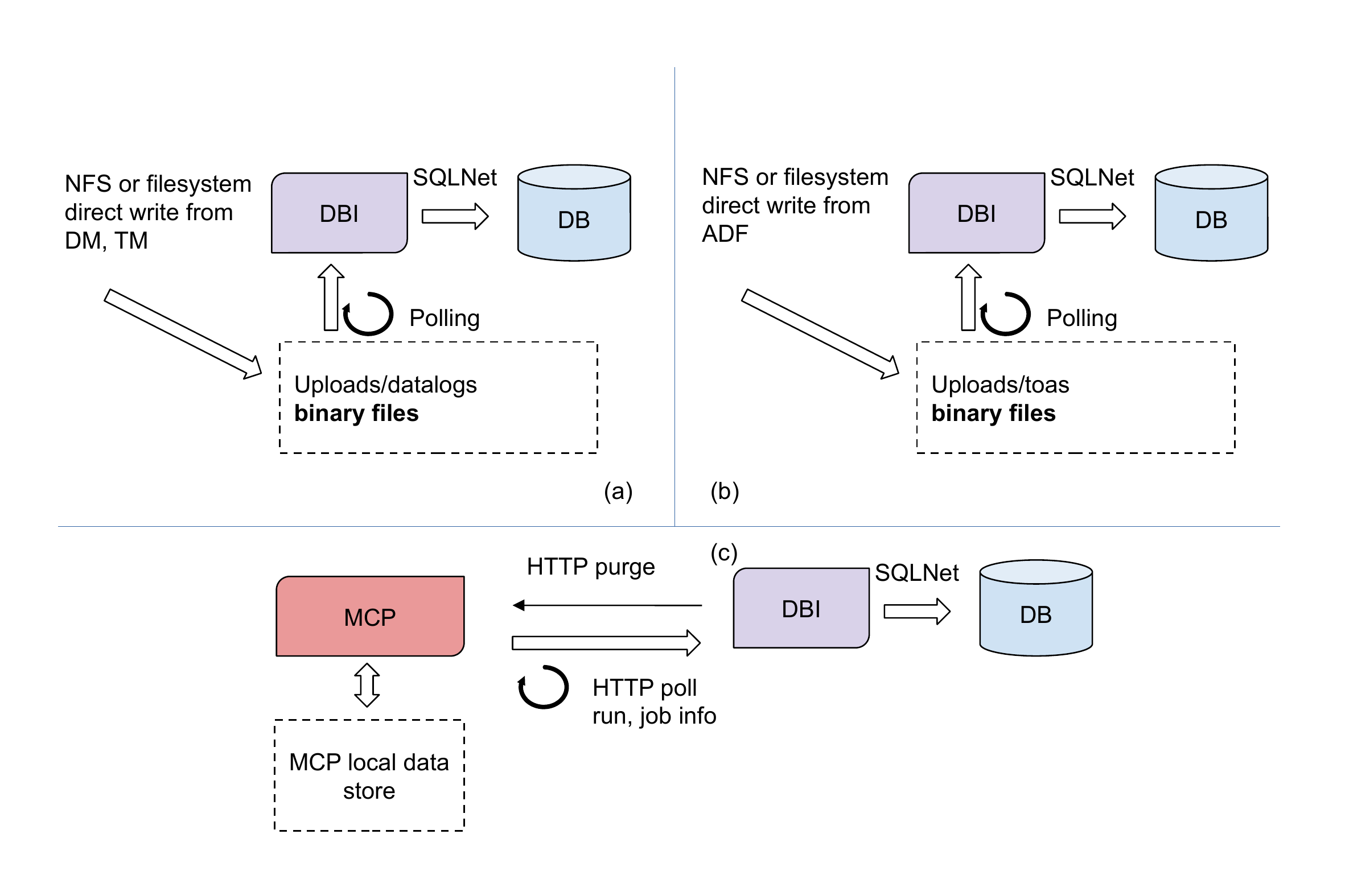}
\caption{\textit{Logic and network protocols involved in data upload to the database. (a) Datalog data flow. (b) TOA data flow. (c) Run and job book-keeping information flow.}}
\label{fig-info}
\end{figure}

Run book-keeping information is ``pulled'' by the DBI querying the MCP and written to the database without going through a local cache. This reflects the fact that \emph{datalog} and TOA tables in the database have foreign keys to the table of runs: an error in \emph{datalogs} or TOAs remains confined to that dataset, but an error in run book-keeping would have a cascade effect of errors on other tables. The DBI will send a ``purge'' command to the MCP for runs and jobs that have been successfully written. \emph{Datalogs} and TOAs for the runs and jobs that have been already communicated to the database and staged in the upload cache are cleared for writing to the database, whereas all other data therein are kept standing by.  Figure \ref{fig-info} shows the different information flows.

When a \emph{datalog} or TOA set write fails, it is not retried until another write of \emph{datalog} or TOA set succeeds. This copes with the case of Wide Area Network failure: for a certain timespan all writes fail, but each dataset is tried only once. As soon as the database can be reached again, all queued writes are executed. If a dataset cannot be written multiple times (usually the threshold is set to 5), it is flagged as ``failed'' and must be reviewed by a DAQ expert.

\section{Detector management - DM}
\label{DM}

\begin{figure}[ht!]
\centering
\includegraphics[width=0.7\textwidth]{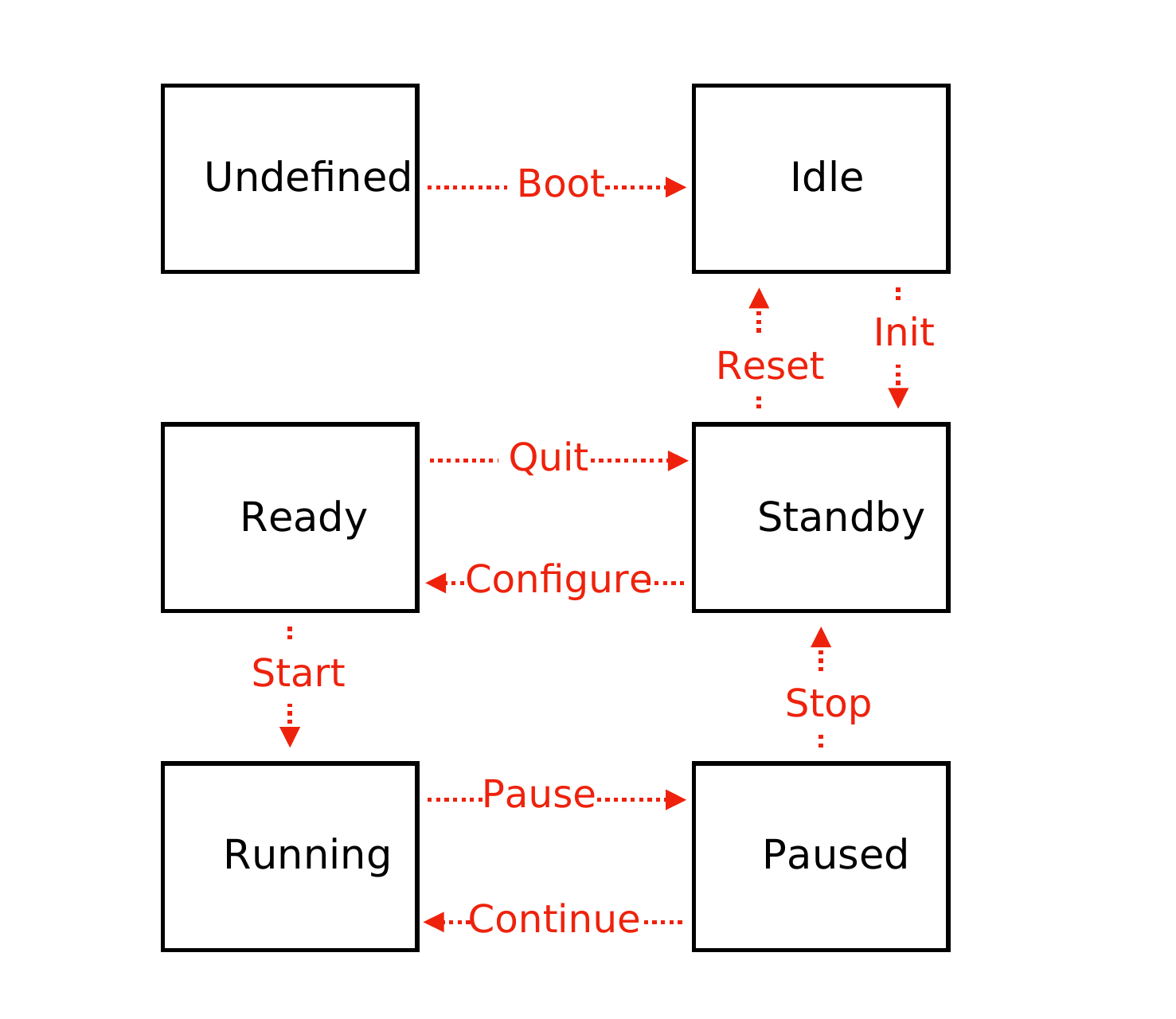}
\caption{\textit{The state machine for the data acquisition as implemented in both the CU and the CLBs: the states are boxed, while the events are paired to the dashed arrows that indicate the related state transitions.}}
\label{fig-statemachine}
\end{figure}

Detector subsystems work according to the state machine depicted in Figure \ref{fig-statemachine}. The three states ``Idle'' (corresponding to the ``Off'' target), ``Ready'' (corresponding to the ``On'' target) and ``Running'' (corresponding to the ``Run'' target) are stable, in the sense that they are supposed to be kept throughout the duration of a run job lasting several hours. The ``Standby'' and ``Pause'' states are transitional. The DM drives the state machine of each CLB issuing events that are transported over the network. The task of the Detector Manager is threefold:

\begin{enumerate}
    \item setting the input parameters of DOMs as specified in the current \emph{runsetup};
    \item driving the state machines of the CLBs according to the current target;
    \item reading out and logging the output parameters of DOMs producing the \emph{datalog} files ready to be written to the database by the DBI (see Sect. \ref{DBI}).
\end{enumerate}

The DM is indeed the most critical component of the CU from the point of view of scalability to the size of a 115-DU block and beyond. It receives messages from all CLBs and sends messages to all CLBs, so it is expected to have CPU and memory footprints that are linearly dependent on the overall number of DOMs.

The DM is expected to receive all notifications from the MCP when the run state changes. As mentioned above, even if the ``push'' mode misses a beat or a communication error occurs, the DM regularly polls the MCP to know the run state. In fact, this allows the DM to work even if the system is being reconfigured while data acquisition goes on. Any run state change will be logged so that it can be written to the database.

Every time the detector or the \emph{runsetup} changes, the DM goes through all DOMs to reconfigure them. This means working out the full list of input parameters and their values and output parameters. Such list is customised to the level of a single PMT. The DM communicates with the CLBs by means of the Simple Retransmission Protocol (SRP - see Subsect. \ref{SRP}), a UDP (User Datagram Protocol)-based protocol. It includes functions to set up and establish the link between the DM server and the CLBs. This is useful both when the DM first starts up, when DUs are rebooted or when a CLB needs to be restarted. One of the purposes of the DM is monitoring the activity of the CLBs and regaining control of those that may stop communicating, thus minimising the need for human interventions. SRP allows point-to-point messages from DM to the CLBs and back, broadcast messages and subscription-based data transmission, so that the DM asks once for the set of parameters to be monitored and receives regular updates (1 Hz or 0.1 Hz) without the need for further polling.
Each CLB exposes the following subsystems:

\begin{itemize}
    \item System (SYS)
    \item Network (NET)
    \item Optics (OPT, only for CLBs hosted in DOMs)
    \item Acoustics (ACS)
    \item Instrumentation (INS)
    \item Base (BSE, only for DU base modules)
\end{itemize}

Each subsystem is controlled independently of the others. However, except during the short timespans of transitions, all subsystems should be in the same state. The DM takes actions when it receives a new CLB status report: it is compared to the currently expected state and, only if they are not in agreement, a new event is generated so the state machine moves to another state. Parameter setting is only allowed in the Configure event that connects the Standby state to the Ready state. Hence, any change in parameters implies driving the CLB state machine to the Standby state, setting the parameters and then putting the state machine in the state that is consistent with the current target. In doing so, also the run number is compared to the corresponding monitoring variable shown by the CLB. If they differ, the DM directs the CLB to go through all the states needed, until the CLB run number matches the current run number defined by the MCP.

For testing and troubleshooting, the DM also provides a manual mode that is reserved to users that hold the \emph{Detector Control} privilege (usually \emph{Run Coordinators} and \emph{DAQ experts}). The manual mode can be activated on single CLBs and allows operators to tweak every single parameter and to control the state machine issuing events manually via a GUI. When the “automatic” control mode is restored, the CLB goes back to normal operation, but newly set input parameters are not restored until the next run switch. The ability to control parameters manually is useful to fix possibly critical conditions while a new \emph{runsetup} is being prepared and a new run is ready to start.

For DU base modules, it is also possible to use the GUI to toggle the DU power. This function is reserved to holders of \emph{Detector Control} privilege. Some parameters can be tuned only through the DM console command line, as they may cause severe damage to the detector, such as overcurrent.

\begin{figure}[ht!]
\includegraphics[width=0.9\textwidth]{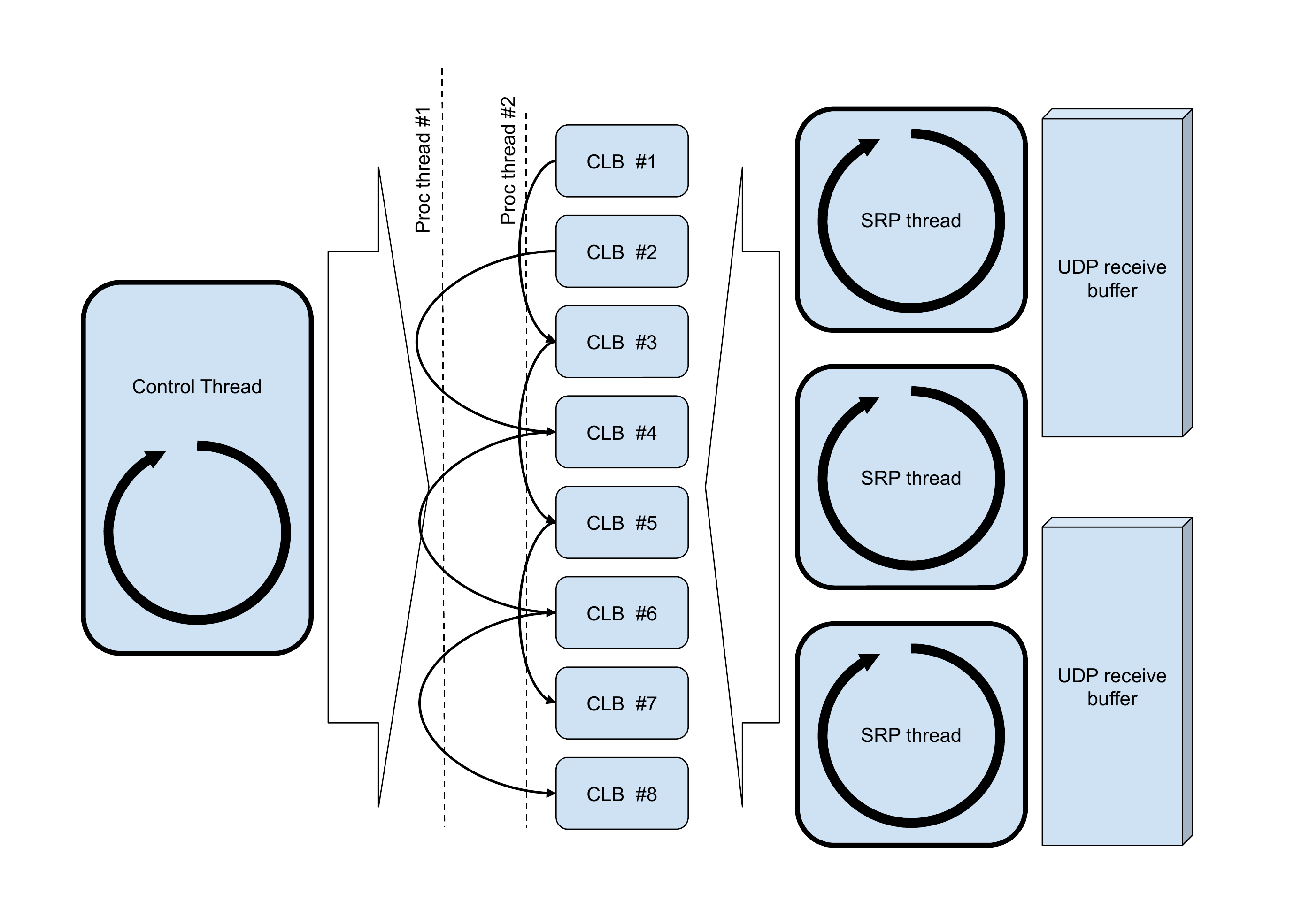}
\caption{\textit{Sketch of the threading structure of the DM. The HTTP thread pool is not shown. The Control Thread sends messages to all CLB Controllers, which have no thread of their own. Processing threads (2 in the sketch) power the CLB Controllers by sharing the workload. SRP threads (3 in the sketch) read the messages found in two UDP socket buffers and convert them into events for CLB Controllers. Large arrows show the communication flow towards CLB Controllers. Small arrows show the sharing of CLB control workload among processing threads.}}
\label{fig-dm}
\end{figure}

As shown in Figure \ref{fig-dm}, the DM has one Control Thread to handle a serial queue of external commands (mostly from the MCP, but also from shifters and console commands by administrators). There is one CLB Controller per CLB, but this does not have its own thread: the usage of computing resources by the DM has to be carefully controlled. Although it is a naturally multi-threaded application, the usage of thread pools is limited to the HTTP interface. The allocation of memory and threads for SRP communications and for CLB action processing is statically configured. It can be changed by explicitly setting configuration parameters in the DM console, but it cannot change during a run. The UDP receive-buffer size can be statically configured. In case of oversubscription, i.e. when too many SRP messages arrive, a fraction of them is automatically dropped. Monitoring messages are grouped by type and source; in case of excess load on the processing thread, subsampling occurs by dropping a suitable fraction of messages. Such loss of information turns out as a decrease of the average sampling frequency of the detector monitoring. The DM provides counters to diagnose the communication and computing load, so that DAQ experts can adjust the allocation of resources. As a reference, sampling a DU at 1 Hz uses about 10\% of one typical CPU core. This implies that about 12 cores should be enough for the monitoring of a whole block of 115 DUs. It has been shown that a single socket with 64 KiB receive-buffer can monitor at least three DUs. The number of sockets can be tuned according to the needs, allowing to scale to a full detector of multiple blocks..

The same program for DM is used in the various KM3NeT environments of detector control, such as shore stations, qualification test benches and development installations. In some cases, specific actions that are normal in other contexts may carry high risks because of peculiarities in legacy hardware. The DM has a standard blacklist of such actions (mostly related to power control functions) that need to be individually allowed. An additional module, called ``Authorization Block'', which is compiled to run on a well-identified machine in a single geographic place, enables those actions that are potentially dangerous. The Authorization Block makes sure that an administrator has explicitly unlocked all permitted functions. A DM without an Authorization Block or with a locked one would filter all actions in the blacklist.

\begin{figure}[ht!]
\centering{\fbox{\includegraphics[width=0.9\textwidth]{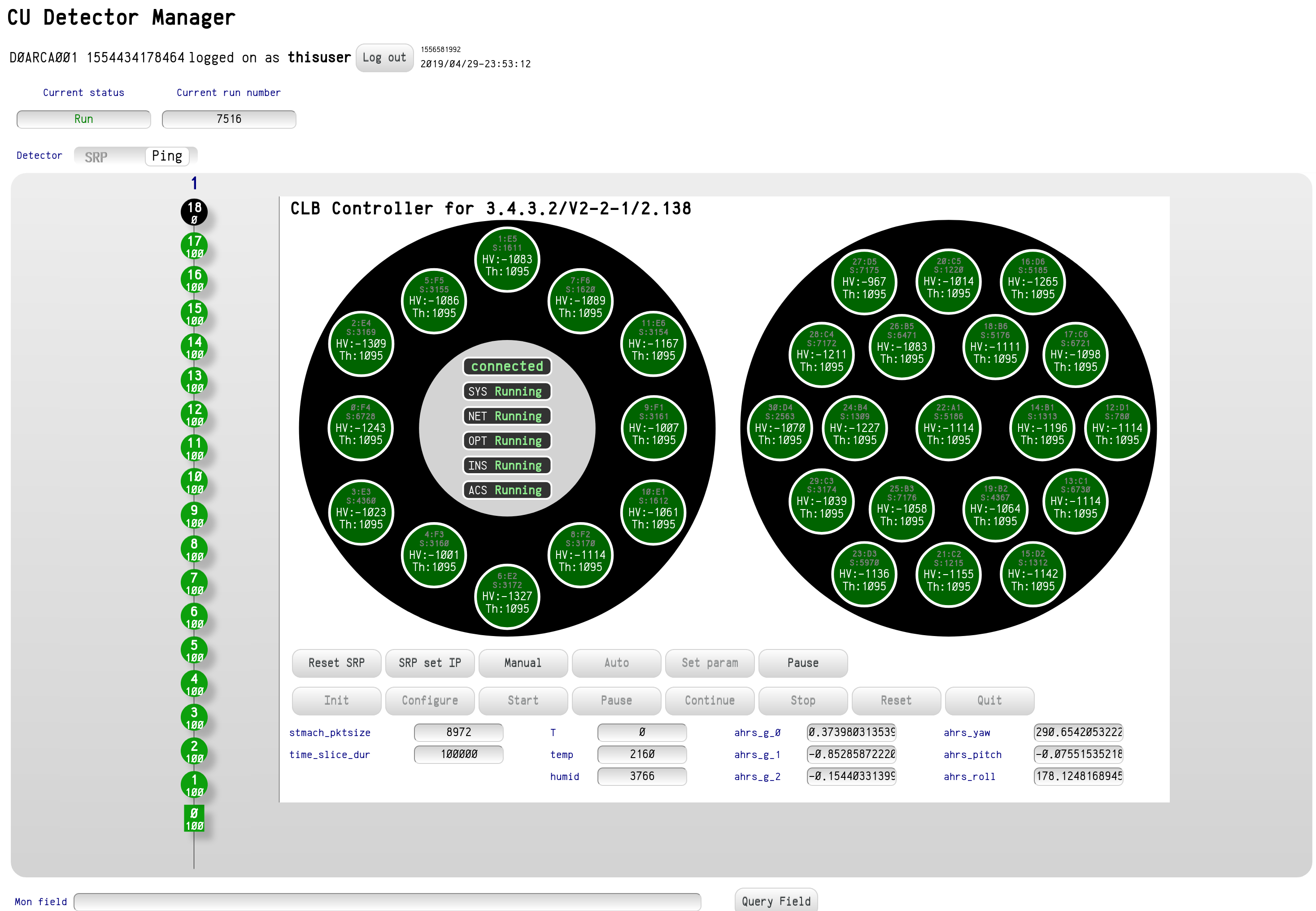}}}
\caption{\textit{Graphical user interface of the DM for one DU and one DOM (superimposed). Live monitoring data obtained via HTTP are shown.}}
\label{fig-dm-screenshot}
\end{figure}

Two outputs are continuously generated by the DM: one is a human-readable log and the other is a binary formatted \emph{datalog}. The latter is produced at regular intervals (usually 10 minutes) or when it reaches a certain size (32 MiB in memory). It contains a chunk of monitoring data ready for database insertion. Usually it is written in the upload cache of the DBI (see Sect. \ref{DBI}). Subsampled snapshots are exposed in the Virtual Directory (see Subsect. \ref{VD}) that is available via the HTTP, mostly for GUI purposes. An example of a screenshot of the GUI with live monitoring data is shown in Figure \ref{fig-dm-screenshot}). In addition, other programs may read them if needed.

\section{TriDAS management - TM}
\label{TM}

\begin{figure}[ht!]
\centering{\includegraphics[width=0.9\textwidth]{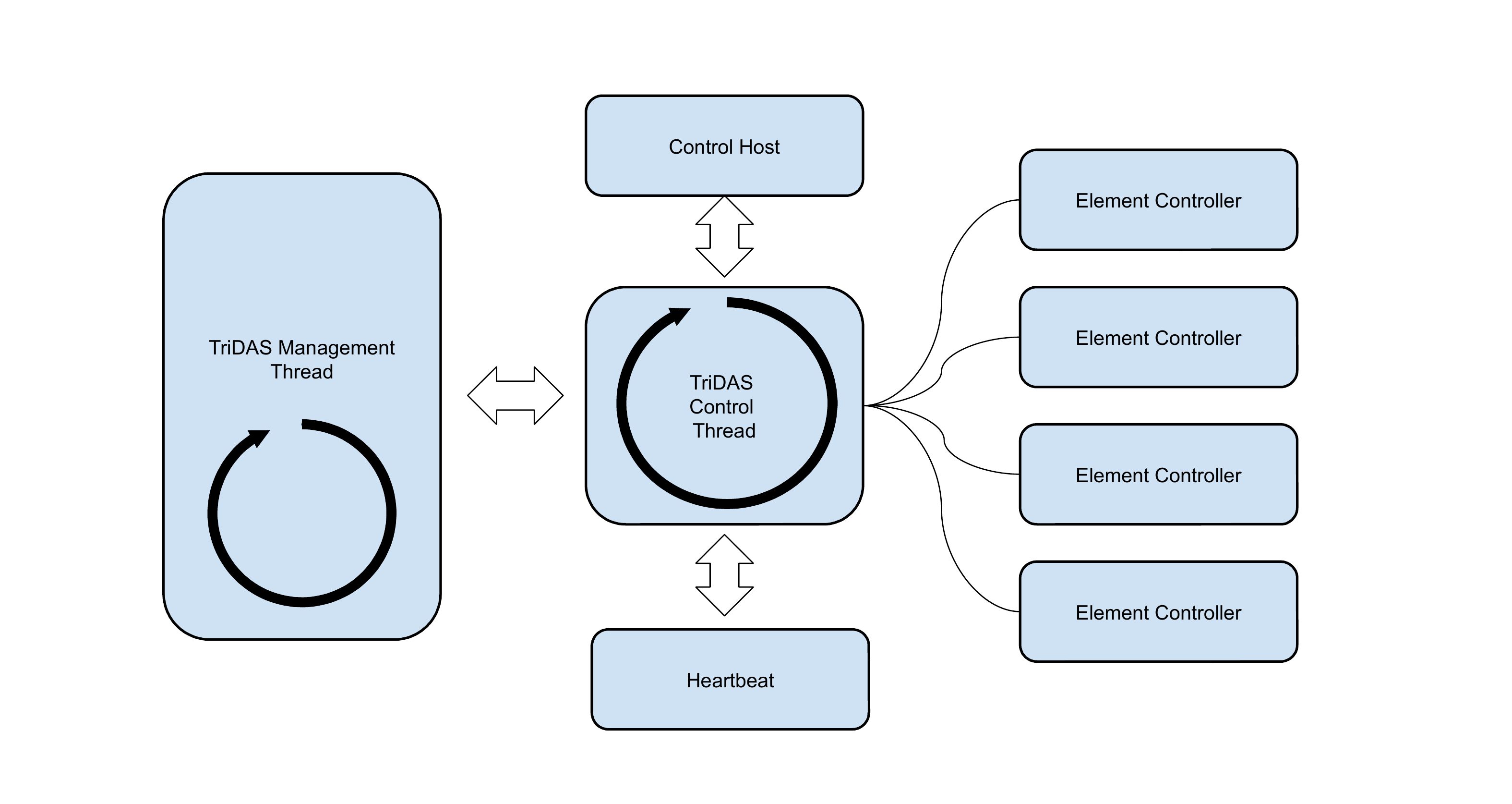}}
\caption{\textit{Sketch of the threading structure of the TM. The TM Control Thread receives commands from HTTP and from the console. The Control Thread powers the Element Controllers while the Control Host interface ensures the I/O and the Heartbeat provides a clock.}}
\label{fig-tm}
\end{figure}

The TriDAS is a set of programs developed in compliance with the requirements of the KM3NeT data taking and processing framework. In most scenarios there is at least one Dispatcher, one or more Optoacoustic Data Queues, one or more Optical Data Filters, one or more Acoustic Data Filters and one or more Data Writers. All the programs need to be driven in a coordinated way, consistently with the current operational target. They all feature a state machine that is identical to the one implemented in the CLBs. As a general guideline, normally all TriDAS components should be in the same state as a generic CLB. Like in the case of the DM, each TriDAS element has its own TriDAS Element Controller. In practice, control and communication are so different for CLBs and TriDAS programs that there are very few similarities in the inner structure of DM and TM. The inner structure of TM is shown in Figure \ref{fig-tm}.

If a CLB suddenly stops responding data taking by the remaining ones goes on unperturbed. In the case of a crash of a TriDAS program, bringing it quickly up again is important to minimise data loss. A crashing ADF is almost harmless if it comes back again within a few hours. An ODF that suddenly disappears leads to a proportional loss (${1 / N_{ODFs}}$) in detector livetime for the duration of the restart procedure. An Optoacoustic Data Queue that crashes leads to total data loss until it is up and running again. Of course, the Data Writer is also crucial because data need to be saved.

While the DM communicates with the CLBs directly one by one, TriDAS processes use the Control Host protocol \footnote{\label{controlhost} Originally developed by R. Gurin and A.Maslennikov (CASPUR, 1995). } to communicate through the Dispatcher. As a result, the TM receives a time-ordered stream of messages from the TriDAS processes. This has some implications on the control process:

\begin{itemize}
    \item The Dispatcher must be identified, contacted and a permanent TCP (Transmission Control Protocol) connection with the TM has to be established.
    \item The Dispatcher cannot be used to start processes (although it can be used to stop them).
    \item While the stream is time-ordered, it is not time-aware, in the sense that it does not produce timeouts like a point-to-point connection does. As a result, a command that is not answered will not automatically produce a timeout error.
\end{itemize}

A local agent (named TriDASManager Agent, or TM Agent) communicates with the TM to receive the requests to start or stop programs and uses the LAP to check that the requests are authorized. The TriDASManager Agent has a security system for credentials that is integrated with the CU. The internal TriDAS Element Controllers have a few more states in their state machine to handle the cases of a program that is starting up, but not responsive yet, or shutting down. A ``Heartbeat'' is introduced to measure time at a central level that is then broadcast to TriDAS Element Controllers.

The TM is a very lightweight application, with a CPU workload that is normally about 3 or 4\% of a single core and only ramps up a bit during run start. It also provides \emph{datalogs} to document starting and stopping times and working conditions for each process.

\section{Networking} \label{Networking}

The CU uses several protocols for communication. This is the result of matching diverse needs and complying with existing standards or practices.

\subsection{HTTP-based access}

All CU services use HTTP as a basic communication protocol. Each CU service uses a lightweight Web-server library that allows exposing an HTTP access. Notice that this is the opposite of what Web-hosting normally means, i.e. hosting the application inside a Web-server. In this case, the application has its own port and its own Web-server dedicated to it. As a matter of fact, a service that runs as a daemon needs a way to communicate with machine administrators, and this often goes through TCP. Using an HTTP interface allows reducing the needs for ports, because both the administrative traffic and user access can go through HTTP.

When applicable, the HTTP server hosts conventional HTML pages for a GUI. They are exposed in the \texttt{/gui} URL directory. Common image formats, CSS style sheets, Javascript source files and AJAX are all supported.

\subsection{SAWI remote calls}

On top of the HTTP layer, the Server Application Web Interface (SAWI) provides a lightweight implementation of remote procedure call. SAWI exposes four virtual directories:

\begin{itemize}
    \item \texttt{/listmethods} gets the list of callable methods;
    \item \texttt{/call} calls a method passing parameters; 
    \item \texttt{/callret} gets the result of a long-running method call;
    \item \texttt{/listcalls} shows the list of calls and their completion status.
\end{itemize}

Subpages of these virtual directories are supposed to be called by programs and have no human-oriented formatting. The pages provided by default at the virtual directories instead show the available options: in practice, a skilled user can mimic remote procedure calls and use the Web browser as a debugger. 

\begin{figure}[ht!]
\centering{\fbox{\includegraphics[width=0.9\textwidth]{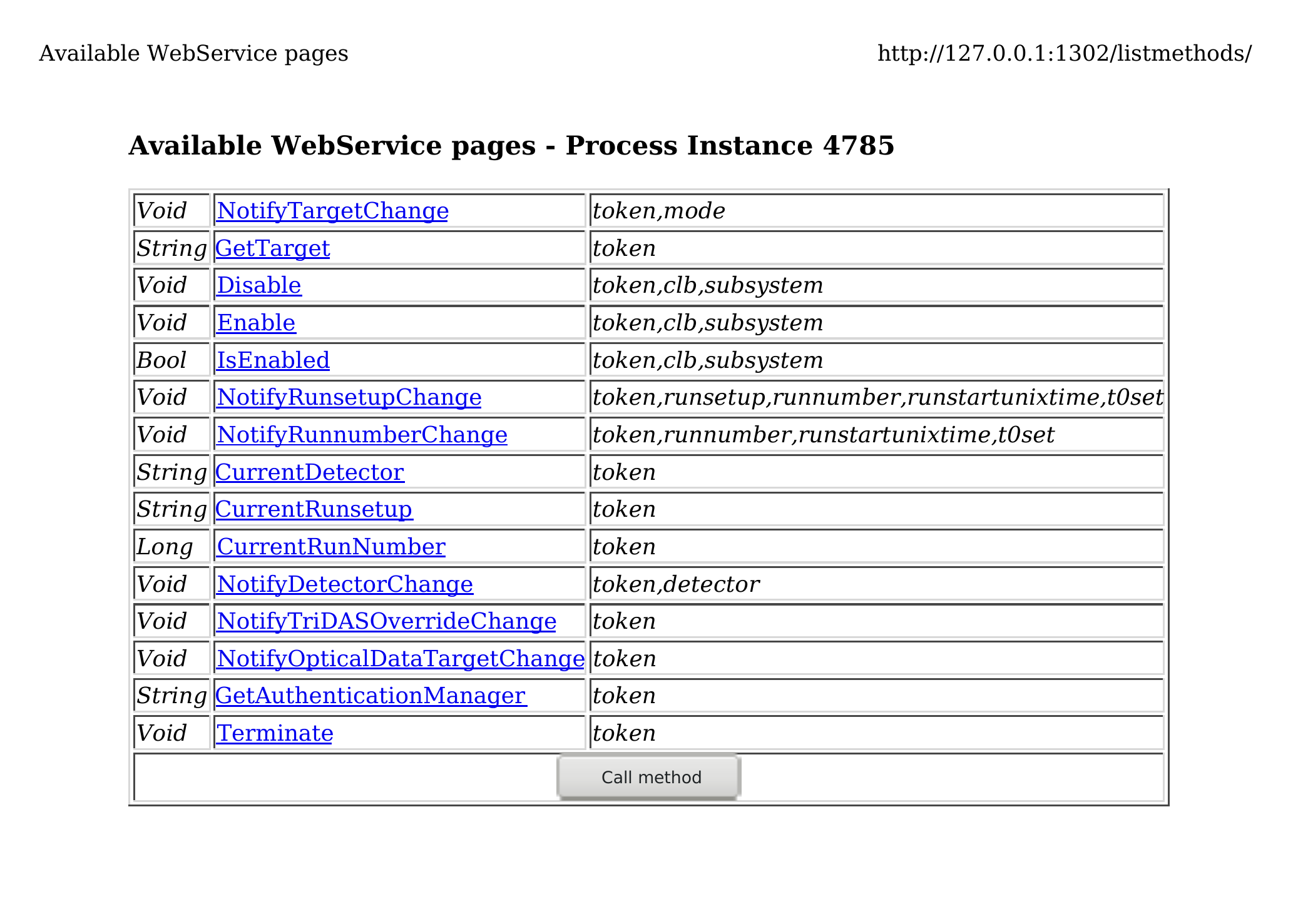}}}
\caption{\textit{SAWI steering page for method calls from a DM. Clicking on each link would show a new page where the arguments to the call can be filled and it can be started. Clients using SAWI would jump directly to subpages, e.g. \texttt{/call/Current\emph{Runsetup}?token=aabbbccc} (``aabbbccc'' is meant to be the security token).}}
\label{fig-sawi}
\end{figure}

SAWI allows both blocking calls and asynchronous calls. The result of an asynchronous call is stored as a job object that is remembered for a set time (usually 10 minutes) after it ends. The caller is expected to poll the \texttt{/callret} virtual directory for completion, specifying the process ID and the ID of the call to get the result. The process ID is provided by the server: in principle, a client has to account for a server process to be restarted, so the call ID is not enough alone to uniquely identify a client-server call. If a server process ID changes, the client knows that the result of the method call is lost and there is no ambiguity. 
SAWI provides support only for simple datatypes (Bool, Int, Long, Double, String). However, it transports exceptions from the callee to the caller and distinguishes exceptions of the remote call protocol from functional exceptions.

From the point of view of the developer, usage of SAWI is very simple: the callee just has to flag the methods that have to be exposed with the \emph{WSrvPage} attribute. The C\# method is reflected at runtime and exposed over the network with the same parameter names. The caller has to include a \emph{WSrvPageClient} object that declares the name and the type of parameters. The first member of all CU calls is a token string that is checked with the LAP to ensure that the caller has the right to call the procedure. Before the call, the server and port have to be set. The \emph{WSrvPageClient} object can be used multiple times. Figure \ref{fig-sawi} shows an example screenshot of the \texttt{/listmethods} steering page for a TriDASManager.

\subsection{Virtual Directory}
\label{VD}

\begin{figure}[ht!]
\centering{\includegraphics[width=0.9\textwidth]{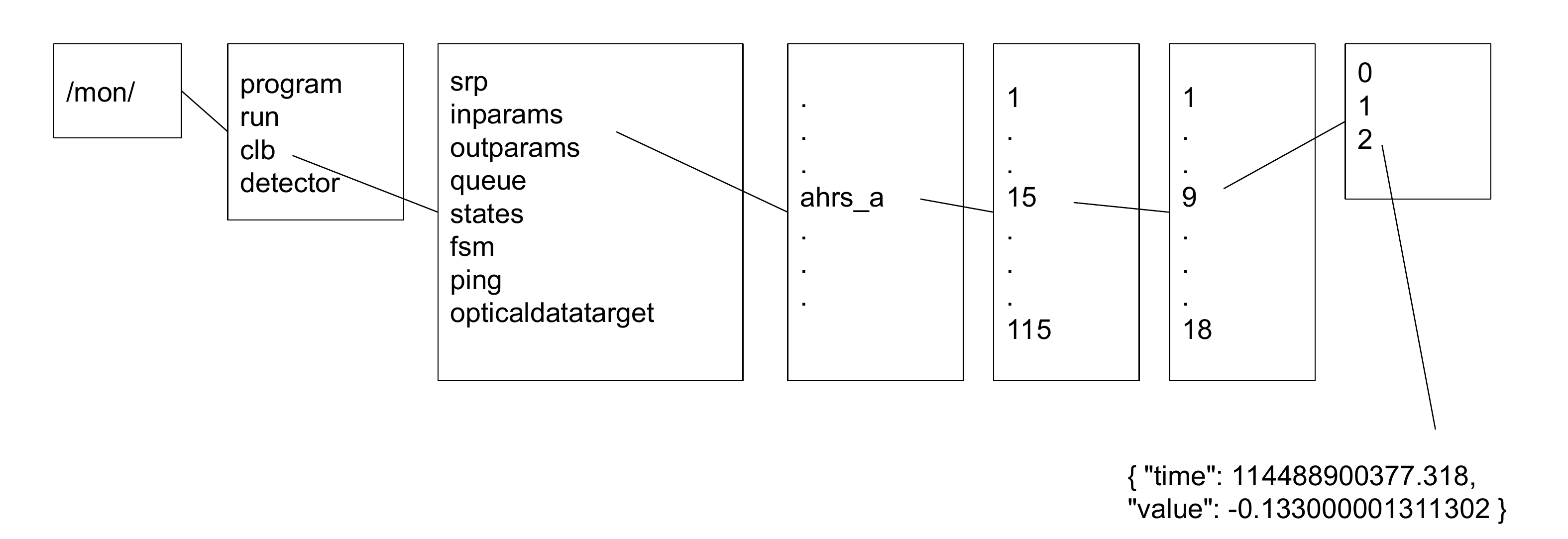}}
\caption{\textit{The KM3NeT detectors change their shape under the action of water currents. The orientation and acceleration of the DOMs are constantly monitored. The example shows the Virtual Directory path to \texttt{/mon/clb/outparams/ahrs\_a/15/9/2} obtaining the vertical acceleration value of the 9\textsuperscript{th} DOM of the 15\textsuperscript{th} DU.}}
\label{fig-vd}
\end{figure}

Each CU service exposes a \texttt{/mon} directory that is meant to contain real-time monitoring data on the service application. They are organized in a virtual directory tree that does not correspond to any file on disk. Leaves in the tree are elementary data, i.e. JSON objects containing the data value and the time it was set. Figure \ref{fig-vd} shows an example of Virtual Directory path to real-time monitoring on the DM.

The implementation of the Virtual Directory structure contains several details that are relevant for optimised performance:

\begin{itemize}
    \item each time a new leaf in the tree is created, the server gets a direct reference to that leaf, which can then be updated without browsing the full path, which would waste CPU power;
    \item HTTP clients are allowed to create shortcuts that gather a client-defined set of variables in a single shot: subsequent calls to the shortcut can retrieve unlimited groups of variables by direct access to their leaves;
    \item writers access direct references to the data leaves in the tree, so they do not need to traverse and lock the tree to repeatedly update values, avoiding mutual locking with readers.
\end{itemize}

Virtual Directory data can be accessed both for the purpose of creating GUI pages or to run monitoring scripts or applications. Using web clients as well as ubiquitous executables such as \texttt{wget} or \texttt{cURL} it is possible to write specific monitors.

\subsection{Simple Retransmission Protocol}
\label{SRP}

The UDP protocol on which the communication between DM and CLBs is based is the SRP. It tags messages and tracks message acknowledgements to allow re-transmission if needed. The DM uses a light version of the SRP library, written in C\#, supporting the subset of the functions that are needed for routine duty. Some diagnostic and debugging features would not be useful in an automatic control context.

\subsection{Control Host interface}

The Control Host library, which is used as the inter-process communication protocol among the data triggering and processing applications, is ported in C\#. The Control Host protocol is used by the TM to connect to the Dispatcher and read / write messages to components of the TriDAS. Each message has a tag and is dispatched to all clients that subscribed for that tag. Since each client can subscribe both for specific tags and for its own unique identifier, both broadcast and one-to-one communication are possible. The Dispatcher collects all incoming messages and enqueues them into serial pipelines. One problem of this protocol is that it requires a persistent TCP connection. Unlike SAWI, which is a connectionless protocol, the Control Host protocol is built for high-speed data transfer but requires a persistent TCP connection. A network error or a disconnection would be interpreted as the client program
closing and reported as such to subscribers.

\section{Dynamic resource provisioning, failover and risk analysis}
\label{DRP}

KM3NeT detectors are expected to operate for at least 10 years. During such timespan, TriDAS servers will be added and upgraded. Some servers will fail and will be replaced with newer ones. Adding, removing and replacing machines should be made easy to help system administrators, who may also change. The importance of maximising the livetime of the detector has already been emphasized. It is worth noticing that it is not only a matter of high percentages of integrated active time, but that even a few consecutive hours of downtime would prevent the observation of rare astrophysical transient events such as supernova neutrino bursts. In this respect, whenever the detector and shore station have enough resources to run, they should be running, even if not at 100\% performance level. This is even more relevant if all powerful servers, which should host ODFs, are in service and just a CU machine has failed. For example, if the machine that hosts the TM fails, the acquisition does not even start, while all the real computing power is there just waiting for a command. A failure analysis has been performed to review the impacts of different failures and assess the corresponding service losses. Conservatively, the mean time between failures of hardware can be estimated to be of the order of five years, but services may be down because of software upgrades, which happens several times in a year and is largely the most common cause of temporary operation interruption, although for short time intervals, of the order of 10 minutes. The analysis is not limited to the CU but also includes the parts of the TriDAS that directly depend on the CU and considers failures caused by one or two concurrent events.

\begin{table}[ht!]
    \begin{tabularx}{\linewidth}{X X X X}
    \hline\\
    \textbf{First condition} & \textbf{Second condition} & \textbf{Loss of service} & \textbf{Impact on data loss} \\
    \hline\\
    LAP down & N/A & GUI inaccessible & LOW \\
    MCP down & N/A & Current run does not end & LOW \\
    DBI down & N/A & \emph{Datalog} + TOA upload pause & LOW \\
    DM down & N/A & Missing \emph{datalogs} & MEDIUM \\
    TM down & N/A & None & LOW \\
    DM down & Run switch & Data loss (run number lag), missing \emph{datalogs} & HIGH \\
    TM down & Run switch & Data loss (run number not set), missing \emph{datalogs} & HIGH \\
    OADQ server down & N/A & Partial data loss & HIGH \\
    OADQ server down & Only available server & Total data loss & HIGH \\
    ODF server down & N/A & Partial or total data loss & HIGH \\
    ODF server down & Only available server & Total data loss & HIGH \\
    ADF server down & Only available server & Total data loss & HIGH \\
    \hline\\
    \end{tabularx}
    \caption{Single-condition and double-condition failure schemes.}
    \label{failures}
\end{table}

As shown in Table \ref{failures}, disentangling different functions and putting them in different programs has already a positive impact on data taking stability, because the first five rows have low or medium severity. Indeed, it is common to upgrade the system during an ongoing run, shutting down services and restarting them one by one. Nevertheless, there are still other high severity scenarios due to multiple failures at the same time. A redundancy in all CU program instances can be introduced by having the same CU service running in multiple machines. This can be obtained with the ``Dynamic Resource Provisioning and Failover'' mode:

\begin{itemize}
    \item The list of machines and services is not defined in the database but it is maintained by the LAP, and continuously logged to the database. This is a natural extension of the basic LAP function of recording users and services. 
    \item For each CU service there are at least two installations, but only one is running while the others are kept in standing by.
    \item When running in Dynamic Resource Provisioning mode, every CU machine runs a LAP that hosts a Health Checker sub-service to perform basic tests. If a Health Checker determines that tests are not passed, it causes the automatic shutdown of the services that are locally hosted: one must avoid that there are conflicting managers, for example two DMs at the same time, one connected to the MCP and the other disconnected.
    \item LAPs poll the Health Checker service to get the status of the machine. A Health Checker answers the status polls that are issued regularly (e.g. 0.1 Hz) in normal conditions. If a machine crashes or fails, its Health Checker does not answer. If the Health Checker answers that the tests are not passed, the machine is not considered suitable to work as if it were failed. The Health Checker itself may fail, but given the fact that the code it runs is very simple, it can be assumed that there is a good (hardware) reason for its failure rather than a bug escaping the pre-deployment testing.
    \item LAPs may reallocate CU services. When they decide to do so, they direct the MCP to switch to another run and the (new) DM and (new) TM to reshape the detector definition and start a new one.
    \item LAPs may reallocate TriDAS computing power. When they decide to do so, they direct the MCP to switch to another run and the DM and TM to reshape the detector definition and start a new one. 
    \item There is no central authority among LAPs. They synchronize their status continuously and services that must exist in single instances are automatically assigned to the available machine with the lowest IP address. Agreement is therefore not imposed by an authority that may itself run on a failed machine, but relies on algorithmic consistency.
\end{itemize}

\begin{figure}[ht!]
\includegraphics[width=0.9\textwidth]{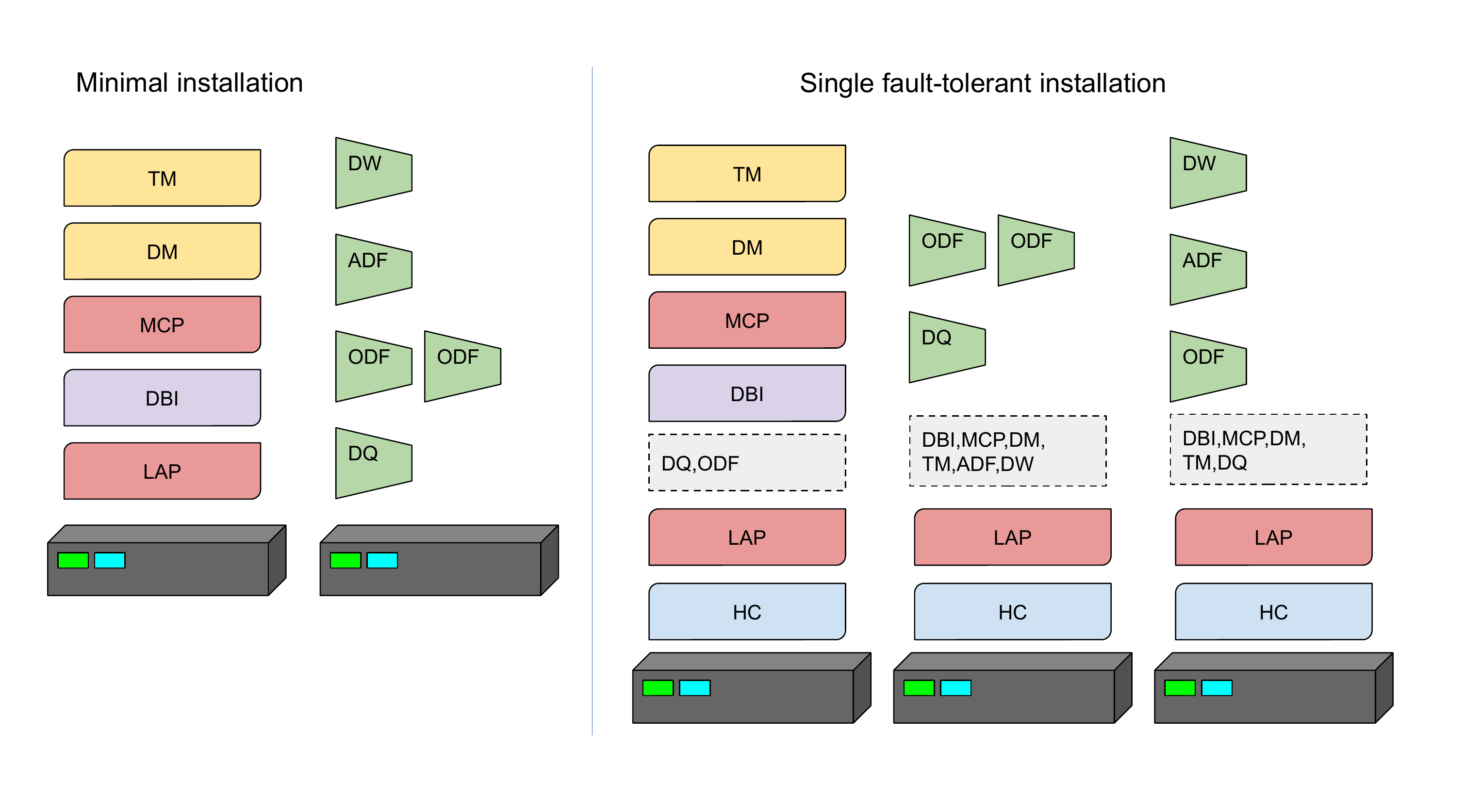}
\caption{\textit{Two configurations of the CU software stack. Left: minimal installation as used in testing stations without fault tolerance. Right: single fault-tolerant installation, as should be used in shore stations. Greyed areas show the services that are installed but kept standing by waiting to take over in case of failures.}}
\label{fig-drp1}
\end{figure}

\begin{figure}[ht!]
\includegraphics[width=0.9\textwidth]{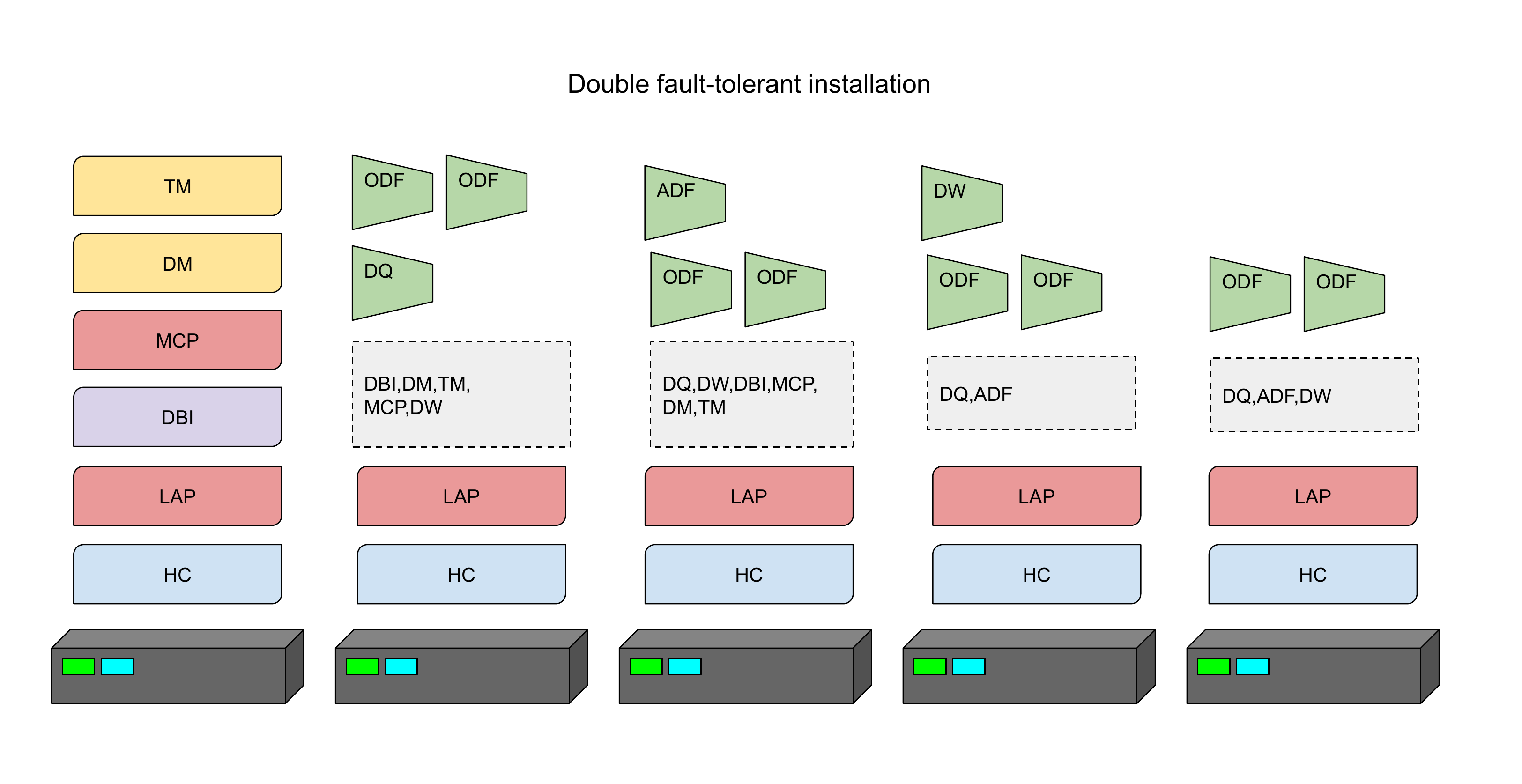}
\caption{\textit{Double fault-tolerant installation: up to two machines may fail at the same time without stopping acquisition, triggering and storage.}}
\label{fig-drp2}
\end{figure}

In this approach, also TriDAS resources are recorded and managed in LAPs, which become a redundant set of local resource managers. The detector definition that MCPs, DMs and TMs get is the current one, i.e. one of all the configurations that are possible with the available resources. In logical compliance to this, at run start, the DM and TM record in \emph{datalogs} the definition of detector that they are using for that run. A detector change always triggers a run change. Figures \ref{fig-drp1} and \ref{fig-drp2} show the full CU stack in various configurations.

The number of Data Queues and Optical Data Filters may change with different processing configurations if computing machines fail. However, it is considered better to run with reduced resources than not running at all. Conversely, in this scenario the addition or replacement of a server is done by just registering the machine change on one of the LAP. It will then propagate the information to others and a detector change will soon allow the newly acquired computing power to enter data taking. Switching from one configuration to another should take place in less than 10 seconds, which is compatible with the duration of most astrophysical transient phenomena.

\section{Summary and conclusions} \label{Conclusions}

The Control Unit of the KM3NeT data acquisition is a system built of several components that work together with the common goals of maximising the live-time and data quality of the operated detectors, both in the deep sea as well as  in component testing / qualification stations. Modularity helps achieving the target of reliability, because several parts of the Control Unit are able to continue their activity despite the temporary unavailability of others. The architecture used, based on the HTTP protocol for interprocess communications, ensures maximum openness of data and algorithms. Graphical user interfaces are provided through common Web technologies. It is possible to access the inner status of Control Unit programs by means of any Web browser. Scalability is guaranteed by performance optimisation and careful design choices. Tests indicate that a single common server with 32 cores and 32 GB RAM  can control a full detector made of two building blocks, with a total of 230 Detection Units. Although the Control Unit continuously reads and writes data to the remote authoritative  database of KM3NeT, possible Internet downtimes are handled without interrupting the detector operations thanks to a dedicated caching system. As the Control Unit is usually accessed through private networks, safety practices are mostly focussed on avoiding mistakes that might affect data quality or detector functionality. In order to achieve that, the Control Unit implements a complete system of privileges for specific operator categories, integrated with the central database. To simplify the administration of the DAQ system and enhance fault tolerance, a Dynamic Resource Provisioning and Failover technology has been developed. It enables the Control Unit to cope even with hardware failures of the hosting servers: all the software services of either the Control Unit  and the  Trigger and Data Acquisition System can be automatically restarted on different  machines, exploiting all  the available computing resources coherently to the prefixed redundancy plan.
The Control Unit is currently in service in more than ten sites, including the two shore stations for the ARCA and ORCA KM3NeT detectors in the Mediterranean Sea and various integration and testing stations. The project benefits of the increasing experience on the detector operations and on the continuous feedbacks from the users. This strategy allows for increasing the operational reliability of the Control Unit and provides with widespread knowledge for the lifetime of the KM3NeT scientific program.

\section*{Acknowledgements}
The authors acknowledge the financial support of the funding agencies:
Agence Nationale de la Recherche (contract ANR-15-CE31-0020),
Centre National de la Recherche Scientifique (CNRS), 
Commission Europ\'eenne (FEDER fund and Marie Curie Program),
Institut Universitaire de France (IUF),
IdEx program and UnivEarthS Labex program at Sorbonne Paris Cit\'e (ANR-10-LABX-0023 and ANR-11-IDEX-0005-02),
Paris \^Ile-de-France Region,
France;
Shota Rustaveli National Science Foundation of Georgia (SRNSFG, FR-18-1268),
Georgia;
Deutsche Forschungsgemeinschaft (DFG),
Germany;
The General Secretariat of Research and Technology (GSRT),
Greece;
Istituto Nazionale di Fisica Nucleare (INFN),
Ministero dell'Istruzione, dell'Universit\`a e della Ricerca (MIUR),
PRIN 2017 program (Grant NAT-NET 2017W4HA7S)
Italy;
Ministry of Higher Education, Scientific Research and Professional Training,
Morocco;
Nederlandse organisatie voor Wetenschappelijk Onderzoek (NWO),
the Netherlands;
The National Science Centre, Poland (2015/18/E/ST2/00758);
National Authority for Scientific Research (ANCS),
Romania;
Ministerio de Ciencia, Innovaci\'{o}n, Investigaci\'{o}n y Universidades (MCIU): Programa Estatal de Generaci\'{o}n de Conocimiento (refs. PGC2018-096663-B-C41, -A-C42, -B-C43, -B-C44) (MCIU/FEDER), Severo Ochoa Centre of Excellence and MultiDark Consolider (MCIU), Junta de Andaluc\'{i}a (ref. SOMM17/6104/UGR), Generalitat Valenciana: Grisol\'{i}a program (ref. GRISOLIA/2018/119), La Caixa Foundation (ref. LCF/BQ/IN17/11620019), EU: MSC program (ref. 713673),
Spain.

\section*{Appendix} \label{Appendix}

    \begin{tabularx}{\linewidth}{X X}
    \hline\\
    Acronym & Meaning \\
    \hline\\
    ADF & Acoustic Data Filter \\
    AJAX & Asynchronous JavaScript and XML \\
    CLB & Central Logic Board \\
    CSS & Cascading Style Sheets \\
    CU & Control Unit \\
    CPU & Central Processing Unit \\
    DAQ & Data Acquisition \\
    DBI & Data Base Interface \\
    DM & Detector Manager \\
    DML & Data Management Language \\
    DOM & Digital Optical Module \\
    DQ & Data Queue \\
    DU & Detection Unit \\
    DW & Data Writer \\
    GUI & Graphical User Interface \\
    HTML & Hypertext Markup Language \\
    HTTP & Hypertext Transfer Protocol \\
    HTTPS & Hypertext Transfer Protocol - Secure \\
    JIT & Just in time \\
    JSON & JavaScript Object Notation \\
    LAP & Local Authentication Provider \\
    MCP & Master Control Program \\
    ODF & Optical Data Filter \\
    PMT & Photomultiplier Tube \\
    SQL & Structured Query Language \\
    SRP & Simple Retransmission Protocol \\
    TCP & Transmission Control Protocol \\
    TOA & Time of arrival \\
    TM & TriDAS Manager \\
    TriDAS & Trigger and Data Acquisition System \\
    UDP & User Datagram Protocol \\
    URL & Uniform Resource Locator \\
    VPN & Virtual Private Network \\
    XML & Extensible Markup Language \\
    \hline\\
    \end{tabularx}

\end{document}